\RequirePackage{lineno}
\documentclass[aps,prc,superscriptaddress, showpacs]{revtex4}

\usepackage{epsfig}
\usepackage{amssymb}
\usepackage{bm}
\usepackage{graphicx,color}
\usepackage{dcolumn}
\usepackage{longtable}
\usepackage{epstopdf}
\usepackage{csquotes}
\usepackage{amsmath}

\begin{document}



\title{Neutron Phase Contrast Imaging of PbWO$_{4}$ Crystals for G Experiment Test Masses Using a Talbot-Lau Neutron Interferometer}
\author {K. T. A. Assumin-Gyimah}
\affiliation{Mississippi State University, Mississippi State, MS 39762, USA}
\author{D. Dutta}
\affiliation{Mississippi State University, Mississippi State, MS 39762, USA}
\author{D. S. Hussey}
\affiliation{National Institute of Standards and Technology, Gaithersburg, MD 20899, USA}
\author {W. M. Snow}
\affiliation{Indiana University/CEEM, 2401 Milo B. Sampson Lane, Bloomington, IN 47408, USA}
\author {C. Langlois}
\affiliation{United Shore Financial Services, Pontiac, MI 48341, USA}
\author {V. Lee}
\affiliation{National Institute of Standards and Technology, Gaithersburg, MD 20899, USA}

\date{\today}

\begin{abstract}

The use of transparent test/source masses can benefit future measurements of Newton's gravitational constant $G$. Such transparent test mass materials can enable nondestructive, quantitative internal density gradient measurements using optical interferometry and allow in-situ optical metrology methods to be realized for the critical distance measurements often needed in a $G$ apparatus. To confirm the sensitivity of such optical interferometry measurements to internal density gradients it is desirable to conduct a check with  a totally independent technique. We present an upper bound on possible internal density gradients in lead tungstate (PbWO$_4$) crystals using a Talbot-Lau neutron interferometer on the Cold Neutron Imaging Facility (CNIF) at NIST. We placed an upper bound on a fractional atomic density gradient in two PbWO$_{4}$ test crystals of ${1 \over N}{dN \over dx}<0.5 \times 10^{-6}$ cm$^{-1}$. This value is about two orders of magnitude smaller than required for $G$ measurements. We discuss the implications of this result and of other nondestructive methods for characterization of internal density inhomogeneties which can be applied to test masses in $G$ experiments.  

\end{abstract}



\maketitle

\section{Introduction}

\begin{figure}[h!]
  \begin{center}
  \includegraphics[width=11cm]{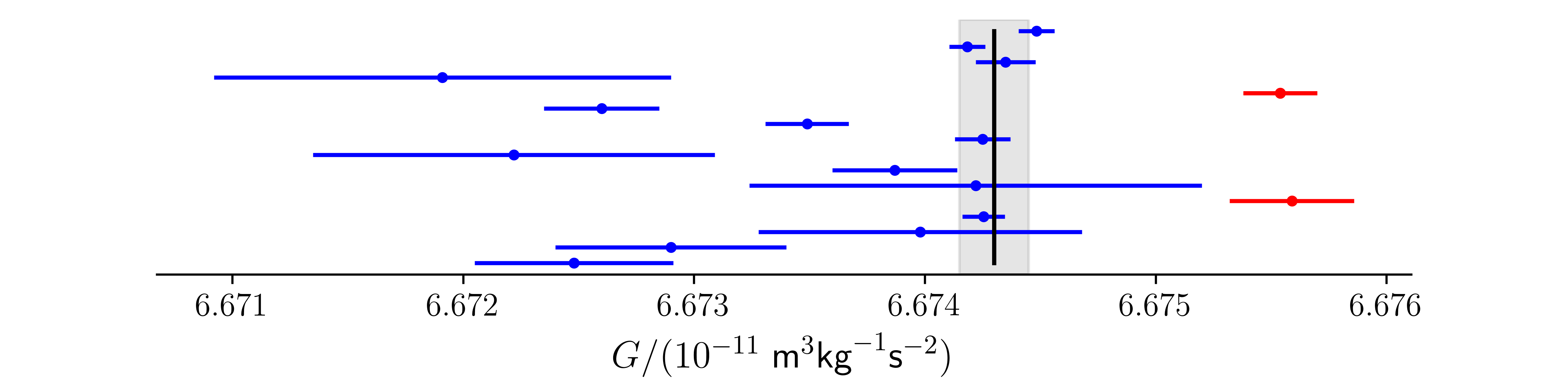}
  \caption{\textit{The most precise measurements of $G$ over the last 30 years. The blue points were used to produce the 2019 CODATA recommended value. The red data points come from two new recent measurements of G~\cite{Li2018}.}}
  \label{fig:fig1}
  \end{center}
\end{figure}

Despite a long history, starting with the 1798 measurement by Henry Cavendish~\cite{cavendish}, our current knowledge
of the universal gravitational constant $G$ is rather poor compared to other fundamental constants. Gravity is the weakest force in nature and it is impossible to shield. Measuring $G$ is a serious precision measurement challenge. The 2018 recommended value by the Committee on Data for Science and Technology (CODATA)~\cite{codata19} has a relative uncertainty of $22 \times 10^{-6}$. Figure~\ref{fig:fig1} shows the most precise measurements conducted in the past 30 years, including both the measurements used in the 2018 CODATA evaluation and two newer measurements~\cite{Li2018}. The scatter among the values from different experiments is far beyond what one would expect based on the quoted errors. 


This implies a lack of complete understanding of either the physics behind gravitation, or the systematics used to perform these measurements, or both. General relativity, the currently accepted description of gravitation, may not be complete. A successful unification of gravitation with quantum mechanics remains elusive~\cite{grreview}. 
Typical $G$ experiments measure small forces,
torques, or accelerations with a relative uncertainty of about $10^{-5}$, and it is fair to say that most people believe that the scatter in this data represents unaccounted-for systematic errors which plague the metrology of small forces. Many of these metrological issues are discussed in a recent review~\cite{Rothleitner2017}.

Given this situation, future precision $G$ measurements will justifiably be held to a
higher standard for their analysis and quantitative characterization of systematic errors.
The procedures for corrections to the raw $G$ data from apparatus
calibration and systematic errors using subsidiary measurements is very specific to the particular measurement apparatus and approach. However, there are some sources of systematic uncertainties that are common to almost all precision measurements of $G$, for example the metrology of the source/test masses used in the experiments~\cite{bigGreview}. A research program which successfully addresses this issue can help improve G measurements. 

We are working to characterize optically transparent masses for $G$ experiments. The use of transparent source/test masses in $G$ experiments enables nondestructive, quantitative internal density gradient measurements using optical interferometry and can help prepare the way for optical metrology methods for the critical distance measurements needed in many $G$ measurements. The density variations of glass and single crystals are generally much smaller than those for metals~\cite{gillies2014} and hence they are likely to be a better choice for source/test masses for experiments which will need improved systematic errors. It is also desirable to use a transparent test mass with a large density.

A typical $G$ measurement instrument involves distances on the order of 50~cm and source mass dimensions on the order of
10~cm. The masses are usually arranged in a pattern which lowers the sensitivity of the gravitational signal to
small shifts $\delta R$ in the location of the true center of mass from the geometrical center of the masses by about two orders of magnitude. Under these typical conditions, we conclude that in order to attain a 1 ppm uncertainty in mass metrology, the internal number density gradients of the test masses must be controlled at the level of about ${1 \over N} {dN \over dx}=10^{-4}$/cm and the absolute precision of the distance measurement to sub-micron precision. It is also valuable to use a test mass with a high density to maximize the gravitational signal strength while also bringing the masses as close as possible to each other consistent with the other experimental constraints. 

We have chosen to characterize density gradients in lead tungstate (PbWO$_{4}$) to the precision required for ppm-precision $G$ measurements. We are performing this characterization using both optical interferometry and neutron interferometry. This paper discusses the results of the neutron interferometric characterization.

\section{Relevant Properties of PbWO$_{4}$}

\begin{figure}[h!]
  \begin{center}
  \includegraphics[width=4.8 in]{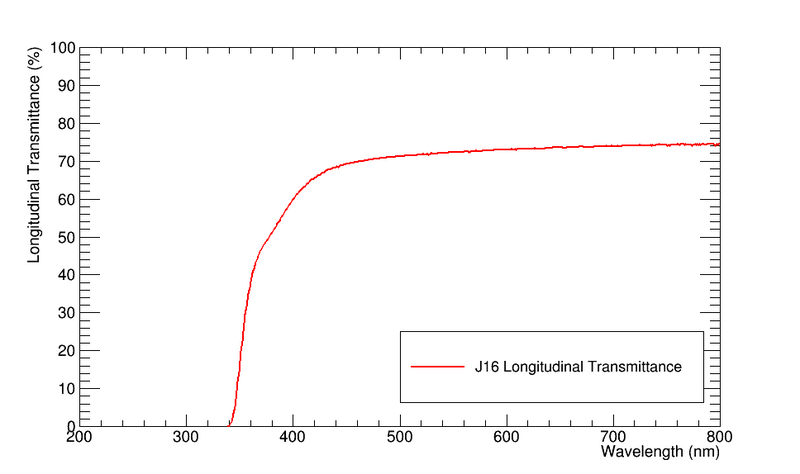}
  \caption{Transmittance of $PbWO_{4}$ as a function of wavelength of incident light.}
  \label{fig:transp}
  \end{center}
\end{figure}

Lead tungstate is a reasonable material to choose for such a characterization. It is dense, optically transparent, nonmagnetic, and machinable to the precision required to determine $G$ and to characterize its internal density gradients by optical techniques. This material can be grown in very large optically transparent single crystals in the range of sizes needed for $G$ experiments and has been developed in high energy physics as a high Z scintillating crystal. The high density uniformity and low impurity concentrations developed to meet the technical requirements for efficient transmission of the internal scintillation light inside these crystals also match the requirements for $G$ test masses. Lead tungstate is transparent for the entire visible spectrum as shown in Fig.~\ref{fig:transp}.

These crystals are non-hygroscopic and are commercially available, for example from the Shanghai Institute of Ceramics (SIC)~\cite{NISTnonendorse}. Several tons of PbWO$_{4}$ are currently in use in nuclear and high energy physics experiments all over the world. This material is actively studied and produced in large quantities for several new detectors under construction and for R\&D on future detectors to be used at the recently-approved Electron Ion Collider to be built at Brookhaven National Lab. We therefore foresee a long-term motivation for continued R\&D on crystal size and quality from nuclear and high energy physics. Based on their scintillation characteristics, the SIC grows boules of PbWO$_{4}$ measuring 34 mm$\times$34 mm$\times$360 mm, which are then diamond cut and polished to the desired size (typically 24 mm$\times$24 mm$\times$260 mm)~\cite{sic1}. SIC has recently produced crystals of 60~mm in diameter and are capable of producing crystals with a diameter of 100~mm and length of 320~mm~\cite{sic2}. These larger sizes are very suitable for $G$ experiments~\cite{decca}. 

Since impurities in PbWO$_{4}$ affect the crystal quality, degrade the optical transmittance properties, reduce scintillation light yield, and produce radiation damage, great effort is put into minimizing or eliminating common impurities such as Mo$^{6+}$, Fe$^{2+}$, Na$^{+}$, K$^{+}$ and Y$^{3}$. Analysis using glow discharge mass spectroscopy (GDMS) indicates that most of the common impurities can be reduced to $<$ 1 ppm by weight~\cite{sic1}. At 32 ppm by weight Y$^{3+}$ is the largest impurity and has a direct effect on the uniformity and scintillation properties of PbWO$_{4}$. Therefore, the distribution of Y$^{3+}$ is  carefully controlled to ensure uniformity in detector grade crystals~\cite{sic1}.

The mass density of PbWO$_{4}$, $\rho= 8.26$ g/cm$^{3}$, is only a factor of 2 smaller than that of tungsten, the densest material commonly used in $G$ measurements. Its transparency opens up the possibility for $G$ experiments to conduct laser interferometric measurements of its dimensions and location, thereby providing a way to cross-check coordinate measuring machine metrology and thereby independently confirm its absolute accuracy. In addition, the existence of such a source mass material might inspire new designs of $G$ apparatus to take advantage of the possibility of in-situ optical metrology to re-optimize the apparatus design tradeoffs between systematic errors and $G$ signal size. In addition to PbWO$_{4}$ there are several additional high density materials that are also optically transparent and could be used as source/test masses. A list of relatively high density materials and their key physical properties is listed in Table~\ref{tab:alt_mat}. Although not listed in the table, silicon could be another choice for a $G$ test mass material in view of the large volumes and very high crystal quality available and developed over decades for the semiconductor industry. Neutron and optical interferometric methods can be applied to all of these materials.

\begin{table}[h!]
  \caption{List of non-hygroscopic, optically-transparent, high-density crystals and their physical properties.}
  \begin{center}
    \begin{tabular}{|l|c|c|c|c|c|} \hline
      Material & PbWO$_{4}$ & CdWO$_{4}$ & LSO & LYSO & BGO \\\hline 
      Density [g/cm$^3$] & 8.3 & 7.9 & 7.4 & 7.3& 7.13 \\
      Atomic numbers & 82, 74, 8 & 48, 74, 8 & 71, 32,8 &71, 39, 32,8 & 83, 32, 8 \\
      Refractive index (light) & 2.2 & 2.2-2.3 & 1.82 & 1.82 & 2.15 \\
      Thermal expansion  & 8.3 (para)& 10.2 & 5 & 5 & 7 \\    
      coefficient(s) [10$^{-6}/^{\circ}$C] & 19.7 (perp)& & & & \\ \hline
    \end{tabular}
  \end{center}  
  \label{tab:alt_mat}
\end{table}  

\section{Use of Talbot-Lau Phase Contrast Imaging for Mass Density Gradient Search}

Slow neutrons can penetrate macroscopic amounts of matter, and their coherent interactions with matter dominate the low-energy dynamics and enable various forms of interferometric measurement methods. We sought a neutron interferometric phase-sensitive measurement method which would be as sensitive as possible to internal density gradients. A neutron passing through a medium of number density $N$ and thickness $L$ will accumulate a phase shift $\Phi=Nb_{c}\lambda L$ where $b_{c}$ is the coherent neutron scattering amplitude and $\lambda$ is the neutron wavelength. The quantity of interest for our study is the fractional density gradient ${1 \over N}{dN \over dx}$. The coherent neutron scattering lengths $b_{c}$ of almost all stable nuclei are known from experiment, and almost all of them possess negligible dispersion so that their values are constant to high accuracy. The neutron wavelength $\lambda$ of slow neutrons can be determined with sufficient precision using a wide variety of methods including diffraction, neutron time-of-flight, etc. Uncertainties in $L$ measurements can be quite small.   

We therefore seek an interferometric technique which can measure a phase gradient ${ \partial \Phi \over \partial x}$. This can be done in principle using different types of phase contrast imaging techniques. Here we briefly review the operation of a specific type of shearing interferometer, a Talbot-Lau interferometer for neutron phase gradient measurement and its application to our problem. We refer readers to the description in~\cite{psinTLI} and to a recent review~\cite{Momose2020} for more details. 

\begin{figure}[hbt!]
    \begin{center}
    \includegraphics[width=11cm]{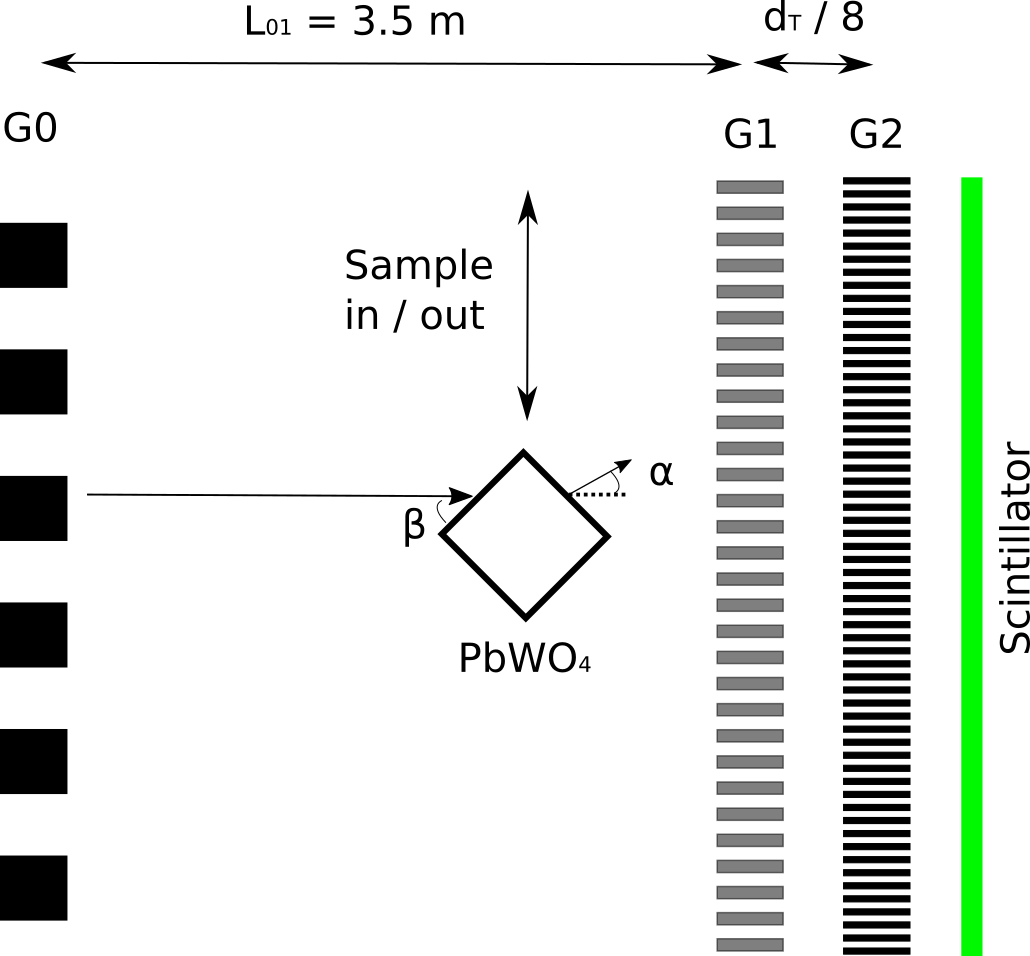}
    \caption{Sketch of the Talbot-Lau phase gradient measurement setup, not to scale, showing the gratings $G_{0}$, $G_{1}$, and $G_{2}$ whose functions are described in the text. The PbWO$_{4}$ crystal is placed so that each face is approximately 45 degrees with respect to the neutron beam axis ($\beta$) which refracts the incident beam by angle $\alpha$. The measurement consists of acquiring phase step images with the sample in and then translated out of the beam. The phase step images were acquired by translating G1 through one period.}
    \label{fig:tlilayout}
    \end{center}
\end{figure}

The Talbot-Lau interferometer sketched in Figure~\ref{fig:tlilayout} consists of three gratings $G_{0}$, $G_{1}$, and $G_{2}$ with their associated periods $p_{i}$. $G_{0}$ is an absorbing grating whose spacing is chosen to convert an incoherent beam of incident neutrons into multiple parallel independent line sources which are mutually incoherent but possess enough transverse phase coherence for Talbot-Lau interferometry. This allows the self-images formed at $G_{2}$ described below to be overlaid constructively and thereby enables phase contrast imaging even with the types of incoherent beams available at neutron sources.

The $G_{1}$ phase modulation grating, chosen in our case to produce $\pi$ phase shifts in the transmitted neutron phase, diffracts the neutron beam and produces a near-field interference pattern (the Talbot-Lau carpet) with maxima at Talbot lengths $d_{T}=p_{d1}^{2}/8n\lambda$ where $p_{d1}$ is the period of $G_{1}$, and $n$ is an odd integer (1 in our case). An absorption grating $G_{2}$ is located at one of the Talbot carpet maxima and generates a Moire pattern that can be recorded on a 2D sensitive neutron detector. This Moire pattern has the form

$$
I(x,y)=A(x,y)+B(x,y) \cos{[{2\pi z \over mp_{d2}}\alpha(x, y) + \Delta (x, y)]}
$$

where $x, y$ are the coordinates transverse to the neutron beam optical axis, which points along the direction of the separation $z$ between gratings $G_{1}$ and $G_{2}$, and $m$ is an integer. A deflection of the neutron beam direction between $G_{1}$ and $G_{2}$, which in our case can be produced by neutron refraction from the sample, induces a corresponding lateral shift of the interference pattern. The neutron refraction angle $\alpha$ can be related to the gradient of the sample phase shift $\Phi(x, y)$ normal to the neutron beam direction

$$
\alpha(x, y)= {\lambda \over 2\pi} {\partial \Phi(x,y) \over \partial x} 
$$

In our experiment the square cross section PbWO$_{4}$ crystals were mounted in the beam so that their faces were oriented at 45 degrees to the neutron beam axis and so that one hypotenuse is centered on the beam. This geometry produces transverse phase gradients and therefore neutron refraction angles of equal magnitude and opposite sign on either side of the neutron beam axis. If the density of the medium is uniform, $\alpha(x, y)$ is independent of $x$. A fractional density gradient ${1 \over N}{dN \over dx}$ in the PbWO$_{4}$ would generate an additional phase gradient ${\partial \Phi(x,y) \over \partial x} = Nb_{c}\lambda L {dN \over dx}$. We can therefore analyze the phase shift image to either search for or place upper bounds on the internal density gradients of interest along a direction normal to the neutron beam.  

To extract the neutron phase shift information of interest from $I(x, y)$ one can measure several phase contrast maps with a grating displaced normal to the neutron beam by different values of $x$. This procedure generates a sinusoidal phase-stepping curve at each $(x, y)$ pixel of the image from which a map of the total phase can be extracted. The contribution to the total phase shift from $\Delta (x, y)$, which comes from imperfections, alignment offsets, etc. in the gratings, can be removed by taking data with the sample absent.



\section{Sample Characterization}

Visual inspection reveals no evidence for any nonuniform local density anomalies from internal voids or inclusions of the type which might introduce systematic errors in $G$ measurements. The detailed shapes of the four 2.3 cm x 2.3 cm area surfaces for each crystal were measured by the NIST Metrology Group on a coordinate measuring machine, with a maximum permissible error (MPE) of $0.3$ $ \mu$m $+ (L/1000)$  $\mu$m where $L$ is in units of mm. The angles between the surface normals for the two pairs of crystal faces to be exposed to the neutron beam varied between $179.93$ degrees to $179.98$ degrees. 

\begin{figure}[hbt!]
  \begin{center}
  \includegraphics[width=11cm]{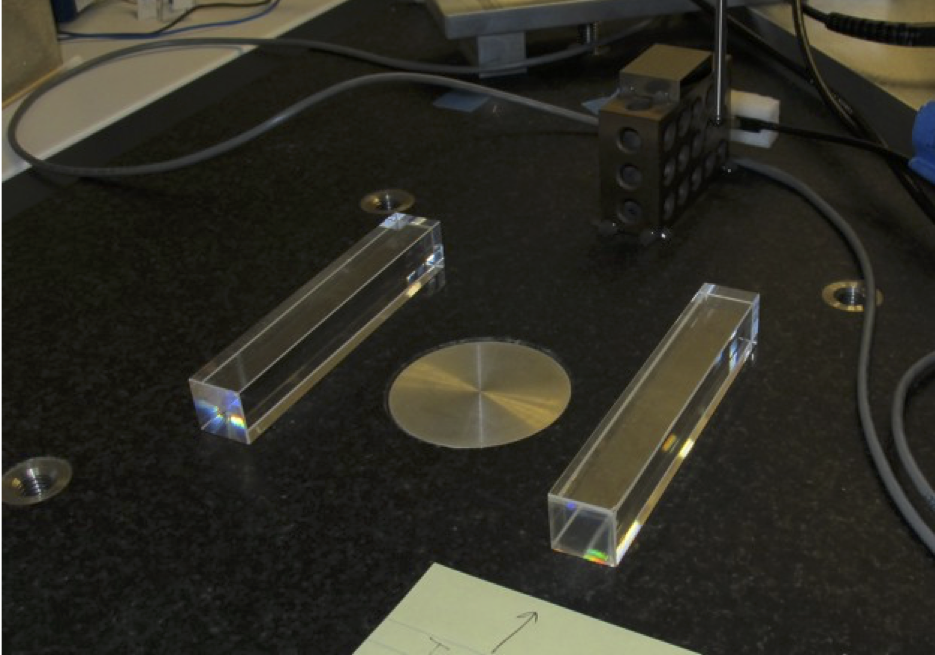}
  \caption{Two PbWO$_{4}$ crystals mounted in the NIST coordinate measuring machine.}
  \label{fig:averagephasegradientplus}
  \end{center}
\end{figure}

\begin{figure}[hbt!]
  \begin{center}
  \includegraphics[width=11cm]{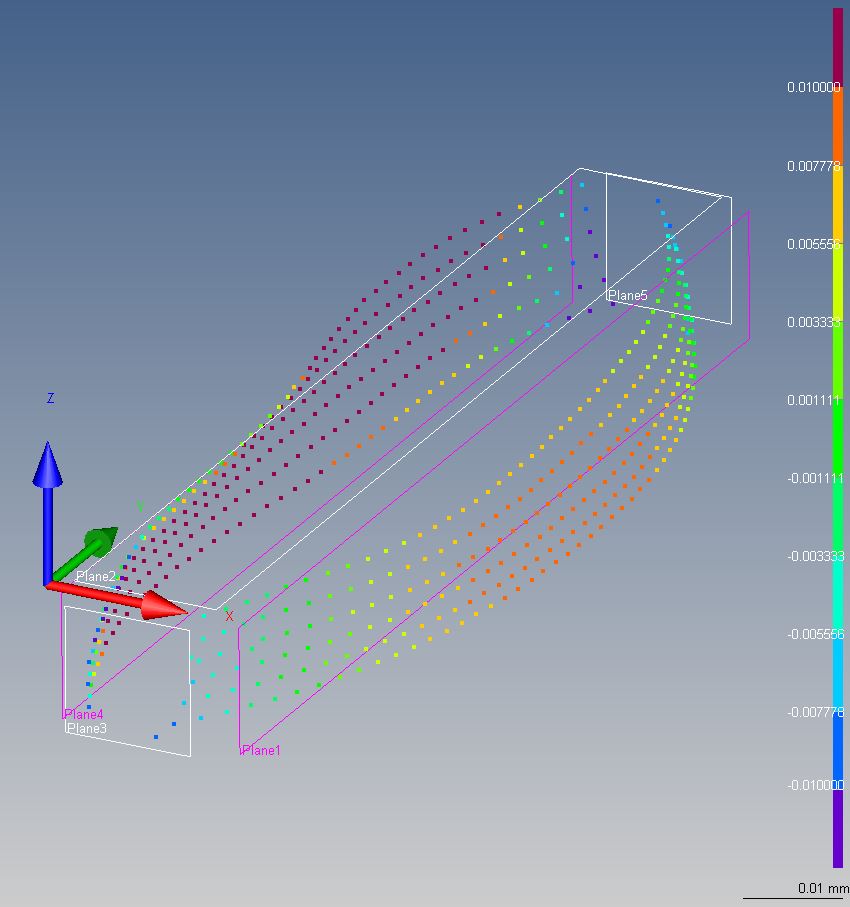}
  \caption{Example map of the surface flatness deviations for two opposing pairs of one of the crystal surfaces as measured by the NIST coordinate measuring machine. The color code on the right extends over a full range of $\pm 0.01$ mm. The size of the measured deviations on the other surfaces of the test masses is similar.}
  \label{fig:averagephasegradientplus}
  \end{center}
\end{figure}
\noindent

\section{Experimental Details}

The density gradients in two nominally identical samples of PbWO$_{4}$ crystals were measured at the NCNR Cold Neutron Imaging Facility (CNIF)~\cite{Hussey2015} at the National Institute of Standards and Technology in Gaithersburg, MD. The PbWO$_{4}$ crystals, both of nominal dimensions 2.3 cm x 2.3 cm x 12 cm, were purchased from the Shanghai Institute of Ceramics in Shanghai, China. The crystal is grown  along the (001) lattice direction. PbWO$_{4}$ has a body-centered tetragonal crystallographic structure with lattice parameters $a=0.54619$ nm and $c=1.2049$ nm (ICDD card n.19-708). 

The neutron beam from the NG-6 cold neutron guide was prepared by diffraction from a double-crystal pyrolytic graphite monochromator and had a mean neutron wavelength of $4.4$ Angstroms. The Talbot-Lau interferometer consists of two vertical gratings G$_{0}$ and G$_{1}$ separated by $3.5$ m, an absorption analyzer grating G$_{2}$ located at the Talbot-Lau self-image location $d_{T}/8$ where $d_{T}$ is the Talbot length, and an imaging detector (described below) in contact with G$_{2}$. The gratings employ Gd coatings to absorb the neutrons. Gadolinium is used for this purpose in view of its extremely high neutron absorption cross section. The initial absorption grating G$_{0}$ consists of 5 $\mu$m thick Gd lines on a 774 $\mu$m period with a duty cycle of 60 \%. G$_{1}$ is a phase modulation grating with 32 $\mu$m deep trenches etched in Si with a period of 7.94 $\mu$m with 50 \% duty cycle. G$_{2}$ is an absorbing analyzer grating with 4 $\mu$m period and 3 $\mu$m high Gd lines.  It is placed about 18.1 mm downstream of G$_{1}$, which corresponds to the $1/8^{th}$ Talbot distance ($7.94^{2}$/($8 \times 0.44$ nm) in close proximity to the scintillator screen. The gratings were made at the NIST Center for Nanoscale Science and Technology nanofabrication facility and are described in the literature ~\cite{Lee2009}. Gratings G$_{0}$ and G$_{1}$ are mounted on 6-axis manipulation stages to enable alignment with the neutron optical axis. The scintillator screen is 150 $\mu$m thick LiF:ZnS scintillator purchased from Applied Scintillator Technologies (now Scintacor). The camera is an Andor NEO sCMOS with 6.5 $\mu$m pixel pitch, viewing the scintillator via a Nikon 50 mm f/1.2 lens for an effective pixel pitch of about 51.35~$\mu$m.

The phase gradient images were obtained by stepping G$_{0}$ through one grating period in 21 equidistant steps in the $x$ direction. The sample was alternately translated into and out of the neutron beam periodically. The median of 3 images was taken to rid the image of spurious background events in the zinc sulfide scintillator induced by gamma rays interacting with the camera from cosmic rays and from the ambient neutron capture gamma background in the neutron guide hall facility. After background subtraction, the amplitude and phase of the signals in each pixel of the image were fit using an analysis procedure described by~\cite{butler2014} and implemented at CNIF in the Matlab analysis environment. 



\begin{figure}[hbt!]
  \begin{center}
  \includegraphics[width=12cm]{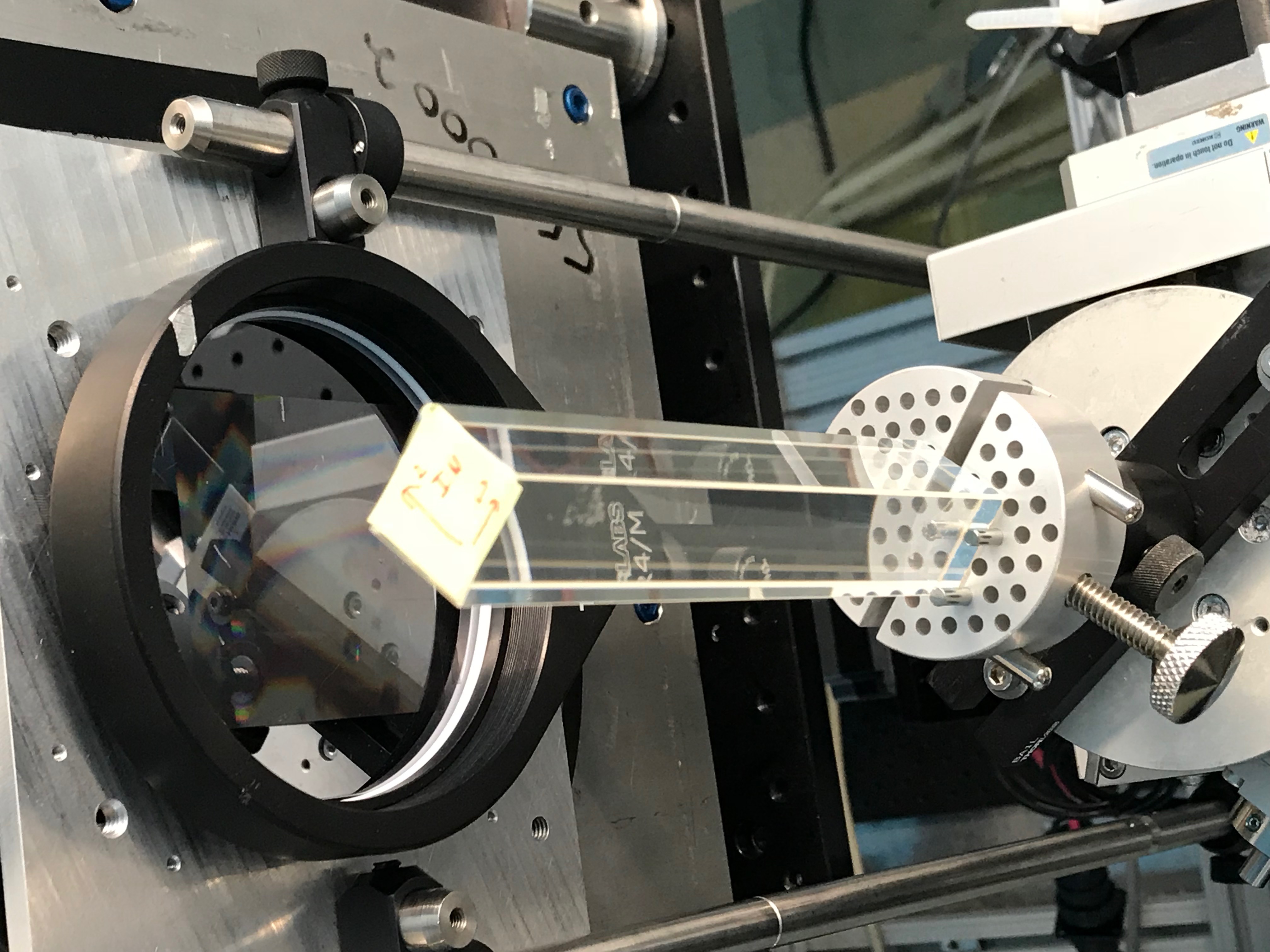}
  \caption{One of the PbWO$_{4}$ crystals mounted in the NIST CNIF beamline. G2 is visible.}
  \label{fig:averagephasegradientplus}
  \end{center}
\end{figure}

For the $i^{th}$ measurement sequence the phase shift is calculated as

$$
\Delta \Phi_{i} = \Phi_{i} - \frac{1}{2} \left(  \Phi_{{\mbox{open}}_{i-1}} + \Phi_{{\mbox{open}}_{i+1}} \right)
$$

\noindent
where $\Phi_{i}$ is the phase shift of the sample for the $i^{th}$ sequence, $\Phi_{\mbox{open}_{i-1}}$ and $\Phi_{\mbox{open}_{i+1}}$ are the phase shifts of the $(i-1)^{th}$ and $(i+1)^{th}$ sequences when the sample was out of the beam. The phase gradient for the sample is found by averaging over the individual calculated gradients.

The fractional uncertainty $\delta (\Phi)$ depends on the number of neutrons registered in the detector, which in turn is the product of the neutron fluence, number of phase steps, height of the region of interest (ROI), the sampling time and the area of a single pixel. The exact values of these critical parameters of the measurement are listed in Table~\ref{tab:params}.

\begin{table}[h!]
  \caption{List of critical parameters}
  \begin{center}
    \begin{tabular}{|l|c|} \hline
Parameter & Value \\ \hline
Pixel pitch & 51.48 $\mu$m  \\
Height of ROI (Sample 1) & 5.148~cm (1000 $\times$ Pixel pitch) \\
Height of ROI (Sample 2) & 5.122~cm (995 $\times$ Pixel pitch) \\
Neutron fluence & 10$^{6}$cm$^{-2}s^{-1}$ \\
Number of phase steps & 21 \\
Sampling time & 45 s \\
Density & 8.30 g/cm$^{-3}$\\
Wavelength($\lambda$) & 4.4 $\AA$ \\
Talbot distance (d) & 7.08 $\times$ 10$^{8} \AA$ \\
Grating period (p$_2$) & 4.0 $\times$ 10$^{4} \AA$ \\
Inclination ($\beta$) & 45$^{\circ}$ \\
Talbot angle ($\theta_{T}$), 1 wedge & 0.7 rad \\\hline
   \end{tabular}
  \end{center}  
  \label{tab:params}
\end{table}

With these parameters $N=1.3 \times$ 10$^{5}$~cm for sample 1 and $1.3 \times$ 10$^{5}$~cm for sample 2. The relative uncertainty is therefore $\delta \phi = 2.8 \times 10^{-3} \phi$ for sample 1 and $ 2.8 \times 10^{-3} \phi$ (the same) for sample 2. The average phase gradient and the average phase gradient uncertainty is calculated as

$$\phi_{AV}^{j} = \sum_{i=1}^{n} \frac{\phi_{i}^{j}}{(\Delta  \phi_{i}^{j})^{2}}; \hspace{5ex} \Delta \phi_{AV}^{j} = \sqrt{ \frac{1}{\sum_{i=1}^{n}\frac{1}{(\Delta  \phi_{i}^{j})^{2}}}},$$ 

where $n$ is the number of measurements to be averaged and $j$ is the number of pixels. 
Using the parameters in Table ~\ref{tab:params} we can calculate the scattering length density as:



$$
Nb_{c} =\frac{p_{2} \theta_{T}}{\lambda^{2} d \tan(\beta)}=2.043 \times 10^{-6} \AA^{-2},
$$
 and the inclination angle of the sample surface from the normal to the neutron beam calculated from the observed phase shift is:

$$
2\theta_{T} =\frac{Nb_{c} \lambda^2 d \tan(\beta)}{p_{2}}= 1.4 rad.
$$

This angle is exactly twice the value in the Table~\ref{tab:params} for the case of a neutron beam incident on a wedge, since the square cross sectional area of the PbWO$_{4}$ sample, when rotated by 45 degrees, presents two wedge surfaces at $\pm 45$ degrees to the neutron beam which both refract the beam into the same direction. 

\section{Analysis}

\begin{figure}[h!]
  \begin{center}
  \includegraphics[width=11cm]{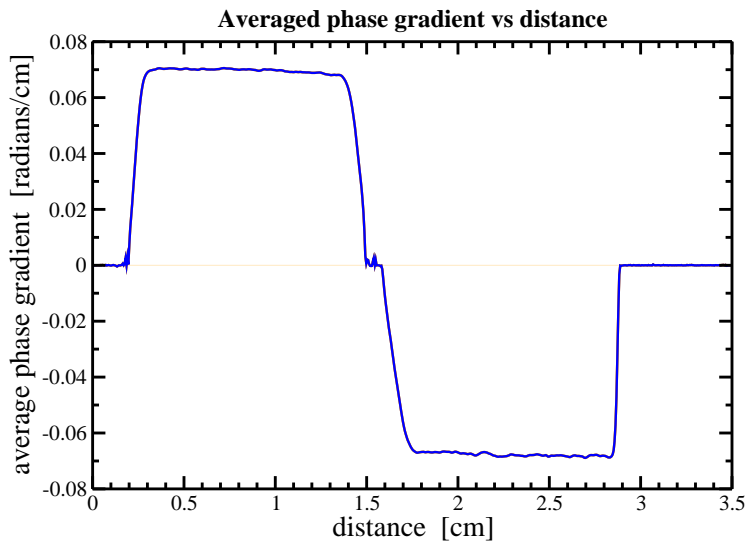}
  \caption{Phase gradient image of the PbWO$_{4}$ sample measured by the Talbot-Lau neutron interferometer.}
  \label{fig:averagephasegradient}
  \end{center}
\end{figure}

Figures ~\ref{fig:averagephasegradient} and ~\ref{fig:averagephasegradientplus} show the averaged phase gradients for one of the PbWO$_{4}$ samples. The opposite signs for $ {\partial \Phi  \over  \partial x}$ on either side of the image are exactly as expected for the 45 degree-inclined square cross sectional area sample employed. These images resolve a very small but nonzero transverse gradient in $ {\partial \Phi  \over \partial x}$ with the same magnitude and sign on both sides. For $0.4 < x < 1.2$ one can calculate the slope of the phase gradient of $d/dx[{\partial \Phi \over \partial x}]=1.9 \times 10^{-3}$ rad/cm$^{2}$. Using the relation $\Phi=Nb_{c}\lambda L$,  differentiating twice, and neglecting possible gradient terms in $N$ and $L$ quadratic in $y$, one picks out the term $2b_{c}\lambda {\partial L \over \partial y}{\partial N \over \partial y}$ of interest in this work. The slope of this phase gradient therefore places an upper bound on the fractional density gradient in the PbWO$_{4}$ of ${1\over N}{dN \over dx}<0.5 \times 10^{-6}$ cm$^{-1}$. The value on the other side of the image is similar. This value is about two orders of magnitude smaller than required for future $G$ measurements. 

\begin{figure}[h!]
  \begin{center}
  \includegraphics[width=11cm]{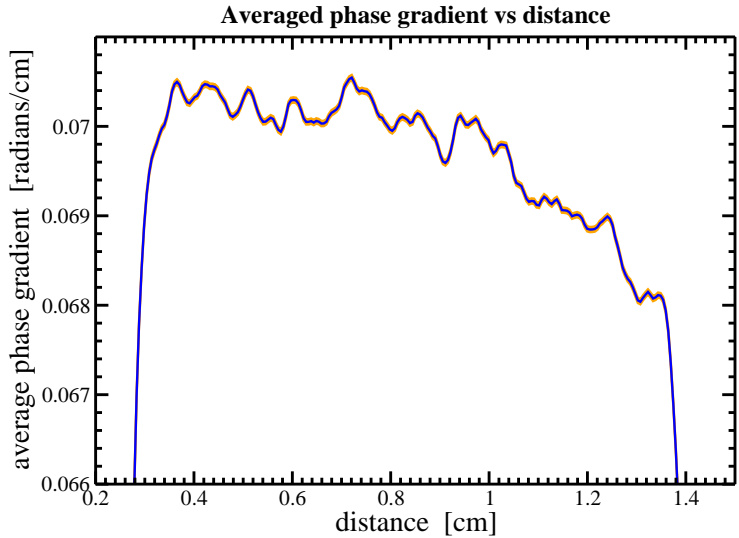}
  \caption{Closeup view of the phase gradient image of the PbWO$_{4}$ sample measured by the Talbot-Lau neutron interferometer. A small but nonzero change in the phase gradient is visible.}
  \label{fig:averagephasegradientplus}
  \end{center}
\end{figure} 
    
\section{Conclusions and Future Work}

We have used a neutron Talbot-Lau interferometer to place an upper bound on the internal density gradient in macroscopic crystals of PbWO$_{4}$ of ${1 \over N}{dN \over dx} < 0.5 \times 10^{-6}$ cm$^{-1}$. This density gradient bound is consistent with the order-of-magnitude one might expect from the level of impurities in commercially-available PbWO$_{4}$ crystals. The size of the phase gradient is also consistent with the NIST coordinate measuring machine measurements of the small deviations of the shape of the crystals away from our assumed geometry.  This density gradient value is about two orders of magnitude smaller than what is needed in typical macroscopic mechanical apparatus of the type often used to measure $G$.

The two PbWO$_{4}$ crystals used in the neutron interferometer studies are currently undergoing testing using a newly constructed laser interferometer. The density gradient will be quantified in terms of variation in the interference fringes as a laser beam scans the crystals. The orientation of the crystal in the optical interferometer in will be kept similar to the orientation in the neutron measurement. If the neutron result is confirmed with the optical interferometer measurements in progress we will conclude that PbWO$_{4}$ is a strong candidate for use as a test mass in future $G$ measurements. 

The same measurement methods used in this work would also work for many other candidate $G$ test mass materials. The neutron measurements of course are not restricted to optically transparent materials and could also be used to nondestructively inspect opaque dense test masses as well. If one wanted an independent confirmation of the results in this case one would need to employ another probe which could penetrate the required thickness of material. This can be done using traditional gamma ray transmission radiography. A gamma ray radiography facility at Los Alamos can routinely penetrate several centimeters of dense material and resolve internal voids at the millimeter scale~\cite{Espy}. It has also recently been shown that an epithermal neutron beamline imaging facility at Los Alamos can at the same time also perform gamma imaging with a comparable spatial resolution. At a spallation neutron source such imaging beams possesses an intense gamma flash when the GeV energy proton beam strikes a high Z target to liberate neutrons by spallation, and the imaging detectors possess both neutron and gamma sensitivity. All three of these methods (neutron and optical phase contrast imaging and gamma transmission) are nondestructive and can therefore in principle be applied to the same $G$ test mass used in the actual experimental measurements. We are therefore encouraged to think that we are close to establishing a method which can address one of the common potential systematic error sources in measurements of $G$.

\section{Acknowledgements}
We would like to thank D. Newell and S. Schlamminger of the National Institute of Standards and Technology in Gaithersburg, MD for their help in arranging the coordinate measuring machine work. K. T. A. Assumin-Gyimah, D. Dutta, and W. M. Snow acknowledge support from US National Science Foundation grant PHY-1707988. W. M. Snow acknowledges support from US National Science Foundation grants PHY-1614545 and PHY-1914405 and the Indiana University Center for Spacetime Symmetries. C. Langlois was supported by the NSF Research Experiences for Undergraduates program. 



\begin{thebibliography}{}
\bibitem{NISTnonendorse} Certain trade names and company products are mentioned in the text or identified in an illustration in order to adequately specify the experimental procedure and equipment used.  In no case does such identification imply recommendation or endorsement by the National Institute of Standards and Technology, nor does it imply that the products are necessarily the best available for the purpose.
\end{thebibliography}
\end{document}